\documentclass[aps,prl,onecolumn,showpacs,preprintnumbers,superscriptaddress]{revtex4}
\usepackage{bm,amsmath,amssymb,latexsym,mathrsfs,enumerate}
\usepackage[mathcal]{euscript}
\usepackage[hypertex]{hyperref}
\usepackage{graphicx}
\usepackage{subfig}

\begin{document}

\title{DYNAMICAL TOROIDAL HOPFIONS IN A FERROMAGNET WITH EASY-AXIS ANISOTROPY}

\author{A. B. Borisov}
\email{Borisov@imp.uran.ru}
\affiliation{Institute of Metal Physics, Ural Division, Russian Academy of Sciences, Yekaterinburg 620041, Russia}

\author{F. N. Rybakov}
\email[Corresponding author. Electronic address: ]{F.N.Rybakov@gmail.com}
\affiliation{Institute of Metal Physics, Ural Division, Russian Academy of Sciences, Yekaterinburg 620041, Russia}

\date{7 September, 2009}
\begin{abstract}
Three-dimensional toroidal precession solitons with a nonzero Hopf index, which uniformly move along the
anisotropy axis in a uniaxial ferromagnet, have been found. The structure and existence region of the solitons
have been numerically determined by solving the Landau–Lifshitz equation.
\end{abstract}


\maketitle

Static and dynamical topological structures with a
nonzero Hopf invariant (hopfions) in various models
and media were discussed in~\cite{bib:Faddeev1,bib:VM,bib:Kamchatnov,bib:DI}. 
In particular, the
magnetization distribution in three-dimensional ferromagnetic materials is characterized by an integer
Hopf index $H$. The investigation of the dynamics of
the hopfions in these media is not only of theoretical
interest, but also important for various physical applications, particularly, in view of the prospect of the 
creation of fundamentally new memory elements.  

Stable precession solitons with $H=0$ (magnon drops) in a uniaxial ferromagnet were found in~\cite{bib:IvKos1,bib:IvKos2}. However, precession uniformly moving toroidal solitons at $H\neq0$ were observed only in an isotropic ferromagnet~\cite{bib:Cooper}. Their stability with respect to perturbations violating the axial symmetry was discussed in~\cite{bib:Sut2007}. 
The aim of this work is to analyze the existence region
and structure of toroidal hopfions uniformly moving
along the anisotropy axis in a uniaxial ferromagnet.

The dynamics of the magnetization vector is
described by the Landau–Lifshitz equation; in the
case of negligible relaxation, it has the form
\begin{equation}
\frac {\partial {\bf M}} {\partial t} = - \gamma \left[ {\bf M} \times {\bf {H_{eff}}}\right],\quad \label{eq:LLeq}
{\bf {H_{eff}} } = - \frac {\delta {E}} {\delta {\bf M}} ,
\end{equation}
where $\gamma$ - is the gyromagnetic ratio ($\gamma>0$). The energy of
the ferromagnet is the sum of the exchange energy
\begin{equation}
E _ {exch}=(\alpha / 2)\int \left({\partial _ {i}} {\bf M} \right)^2 d{\bf r} \label{eq:ExchEnergy}
\end{equation}
and the energy of the uniaxial magnetic anisotropy
\begin{equation}
E _ {anis}=(\beta / 2) \int \left({M _ x}^2+ {M _ y}^2\right) d{\bf r}.\label{eq:AnizEnergy}
\end{equation}
Many magnetic media are characterized by a high
quality factor $Q=\beta/(4\pi)$ and the contribution of the energy of the magnetic dipole interactions is insignificant for them.

At each point of the ${\mathbb{R}}^3$ space with a Cartesian
coordinate system $(x,y,z)$, the orientation of vector $\bf M$ is specified by a point on the two-dimensional sphere ${\mathbb{S}}^2$ in terms of the angular variables $\Theta$ and $\Phi$:
\begin{equation}
M _ x + i M _ y = M _ 0 sin\Theta e^{i \Phi},\quad M _ {z} = M _ 0 cos\Theta .
\label{eq:Mparam}
\end{equation}
Let us consider localized solutions~(\ref{eq:LLeq}), for which
\begin{equation}
\Phi=\omega t + Q \varphi + \phi(r,z - V t),\quad \Theta=\theta(r,z - V t),
\label{eq:Qvortex}
\end{equation}
where $Q\in{\mathbb{Z}}$, $\varphi$ - is the polar angle of the cylindrical
coordinate system $(r,\varphi,z)$, and
$\theta\rightarrow0$ at $|{\bf r}|\rightarrow \infty$. Such solutions describe three-dimensional precession
solitons with a stationary profile, which propagate
along the anisotropy axis. For simplicity, we consider
the configuration of a unit vector field
\begin{equation}
{\bf {n}} = {\bf {n}}(r, z) = (sin \theta cos \phi, sin \theta sin \phi, cos \theta ), \label{eq:nParam}
\end{equation}
in the coordinate system moving along the $z$ axis (with the velocity $V$). 

The desired vector field $\bf M$ specifies the mapping ${\mathbb{R}}^3\cup\{\infty\}\rightarrow{\mathbb{S}}^2$ and is characterized by an integer Hopf topological index $H$. 
If $Q\neq{0}$, the solution given by~(\ref{eq:Qvortex}) corresponds to a toroidal hopfion~\cite{bib:KUR,bib:Glad} with the index
\begin{equation}
H=Q T,\quad T=\frac {1} {4 \pi}\int_{-\infty}^{\infty} {\int_{0}^{\infty} {\bf n}\cdot \left[ {\partial _ {r}} {\bf {n}} \times {\partial _ {z}} {\bf {n}} \right] d{r}}d{z}. \label{eq:HopfIndex}
\end{equation}

Let us represent the energy as a functional of the vector
field $\bf n$:
\begin{equation}
E=\int_{-\infty}^{\infty} \int_{0}^{\infty} w _ E  d{r}d{z},\label{eq:EnergyFunct}
\end{equation}
\begin{equation}
w _ E ={\alpha {M _ 0}^2 \pi r}\left[{(\partial _ {r}} {\bf {n}})^2 + ({\partial _ {z}} {\bf {n}})^2  +\left( \frac {Q^2} {r^2}+{\frac \beta \alpha}\right) ({n _ x}^2 + {n _ y}^2)\right] .\label{eq:sigmaE}
\end{equation}
In addition to energy, equation~(\ref{eq:LLeq}) has two integrals of
motion: the number of spin deviations (magnons),
\begin{equation}
N=\frac {M _ 0} { \gamma  \hbar}\int_{-\infty}^{\infty} \int_{0}^{\infty} (1 - {n _ z}) 2\pi r d{r}d{z}\label{eq:Nintegral}
\end{equation}
and the projection of the momentum of the magnetization field~\cite{bib:PT} on the anisotropy axis,
\begin{equation}
P=-\frac {M _ 0} {\gamma}\int_{-\infty}^{\infty} \int_{0}^{\infty}{\bf n}\cdot \left[ {\partial _ {r}} {\bf {n}} \times {\partial _ {z}} {\bf {n}} \right] \pi r^2 d{r}d{z}.
\label{eq:Pintegral}
\end{equation}

To determine the structure of three-dimensional solitons, we used the same method for minimizing energy
functional~(\ref{eq:EnergyFunct}) with constraint~(\ref{eq:Nintegral}) as in~\cite{bib:BorRyb1}, but constraint~(\ref{eq:Pintegral}) was taken into account through an additional additive square penalty function, and an initial
field configuration was specified by smooth functions $\phi^{}_{ini}$ and $\theta^{}_{ini}$, corresponding to the Hopf bundle with the index $H=Q$:        
\begin{align}
\phi^{}_{ini}&= -\frac{\pi}{2} + v + k \frac{z}{a},\label{eq:fiIni1}\\
cos(\theta^{}_{ini})&= 1-\frac{4 sinh(u)^2 }{2+cosh(2u)}e^{6(1-coth(u))},\label{eq:teIni1}
\end{align}
where $u$ and $v$ are the toroidal coordinates specified by
the relations:
\begin{align}
tanh(u) &= (2 a r)/(a^2 + r^2 + z^2),\quad 0\leqslant u < \infty, \\
cot(v) &= (a^2 - r^2 - z^2)/(2 a z),\quad 0\leqslant v < 2\pi.
\end{align}
The parameters $a$ and $k$ are determined by the integrals of motion $N$ and $P$:
\begin{equation}
a=\sqrt[3]{\frac{3}{4 \pi^2} \frac{\gamma  \hbar}{M _ 0} N},\quad k=k_0 + k_P \frac{\gamma}{M _ 0} \frac{P}{a^2},
\end{equation}
where $k_0\approx{3.353}$ and $k_P\approx{0.076}$.

For comparison, the calculations were performed
by the same method for the case of nontopological
solitons with $H=0$, which are stationary~\cite{bib:IvKos1}
and moving~\cite{bib:Sut2001} magnon drops. The initial field configuration
in these cases was specified by the different functions
\begin{align}
\phi^{}_{ini}&= -\frac{\pi}{2} + k\>atan \left( \frac{z}{r} \right),\label{eq:fiIni0}\\
\theta^{}_{ini}&= \left\{ \begin{array}{l} {\pi \rho/R_1,\quad 0\leqslant\rho<R_1}, \\ {\pi (2 R_1 - \rho)/R_1,\quad R_1\leqslant\rho<2 R_1}, \\ {0, \quad \rho\geqslant 2 R_1}, \end{array} \right. \label{eq:teIni0}
\end{align}
where $\rho=\sqrt{r^2+z^2}$, and the parameters $k$ and $R_1$ are
expressed in terms of the integrals of motion $N$ and $P$:
\begin{align}
k & = 2\frac{\gamma}{M _ 0}P {\left( \frac{\gamma  \hbar}{M _ 0} N \right)^{-\frac{2}{3}}} {\left(\frac{4\pi^2 - 6}{3 \pi^4}\right)^{\frac{2}{3}}},
\\
R_1 & = \frac{1}{2}{\left( \frac{\gamma  \hbar}{M _ 0} N \right)^{\frac{1}{3}}} {\left(\frac{3\pi}{4\pi^2 - 6}\right)^{\frac{1}{3}}}.
\end{align}
The calculations of the nontopological solitons are in
agreement with the data reported in~\cite{bib:IvKos2},~\cite{bib:Sut2001}. 

The desired field ${\bf n}(r,z)$ ensures an extremum of the
functional
\begin{equation}
J({\bf n}(r,z)) = E - \hbar \omega N + V P.
\end{equation}
Using the necessary condition of the extremum, we
arrive at a pair of equations
\begin{align}
\dfrac{d}{d\lambda} J({\bf n}(r + \lambda r,z)) \Bigr|_{\lambda=0} &= 0,\\
\dfrac{d}{d\lambda} J({\bf n}(r,z + \lambda z)) \Bigr|_{\lambda=0} &= 0.
\end{align}
Solving this system with respect to $\omega$ and $V$,
we obtain the convenient formulas:
\begin{equation}
\omega = \frac{E _ {exch} + E _ {anis} - I _ z}{\hbar N }, \quad V = \frac{2 E _ {exch} - 3 I _ z}{ 2 P },\label{eq:omegaV}
\end{equation}
where
\begin{equation}
I _ z = \alpha {M_0}^2 \int_{-\infty}^{\infty} \int_{0}^{\infty}({\partial _ {z}} {\bf {n}})^2 2 \pi r d{r}d{z}. 
\end{equation}

After several thousands of iterations, minimized
energy functional~(\ref{eq:EnergyFunct}) reaches a minimum. To test the
resulting field configurations, we calculated $\omega$ and $V$ by equations~(\ref{eq:omegaV}) and, then, the discrepancy for Landau–
Lifshitz differential equations~(\ref{eq:LLeq}). Let us discuss the results.

\begin{figure}
\begin{center}
\subfloat[]{\label{F:fig1a}\includegraphics[width=0.46\columnwidth, viewport=100 490 330 730, clip]{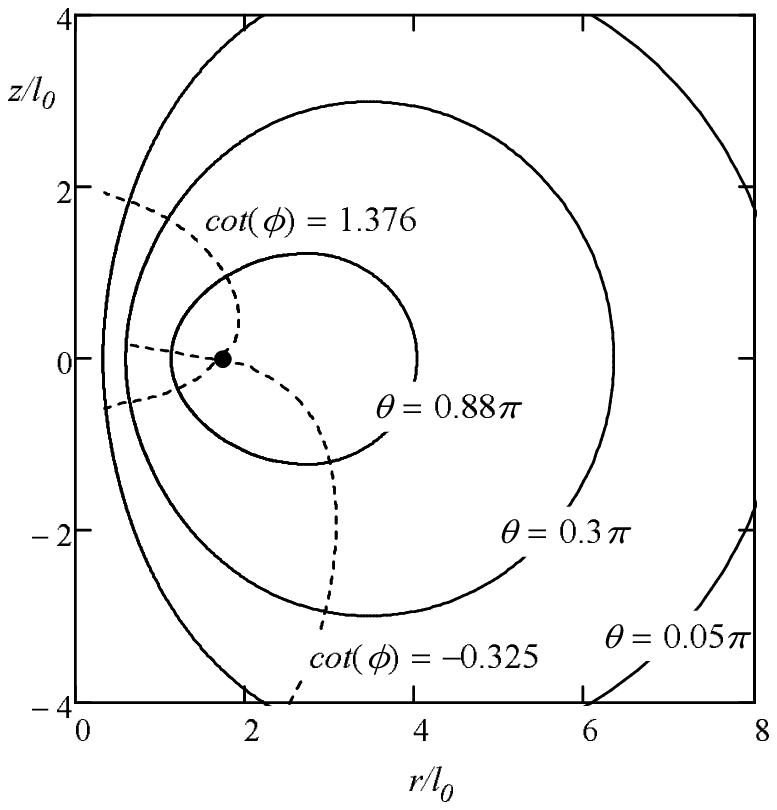}}
\subfloat[]{\label{F:fig1b}\includegraphics[width=0.46\columnwidth, viewport=100 490 330 730, clip]{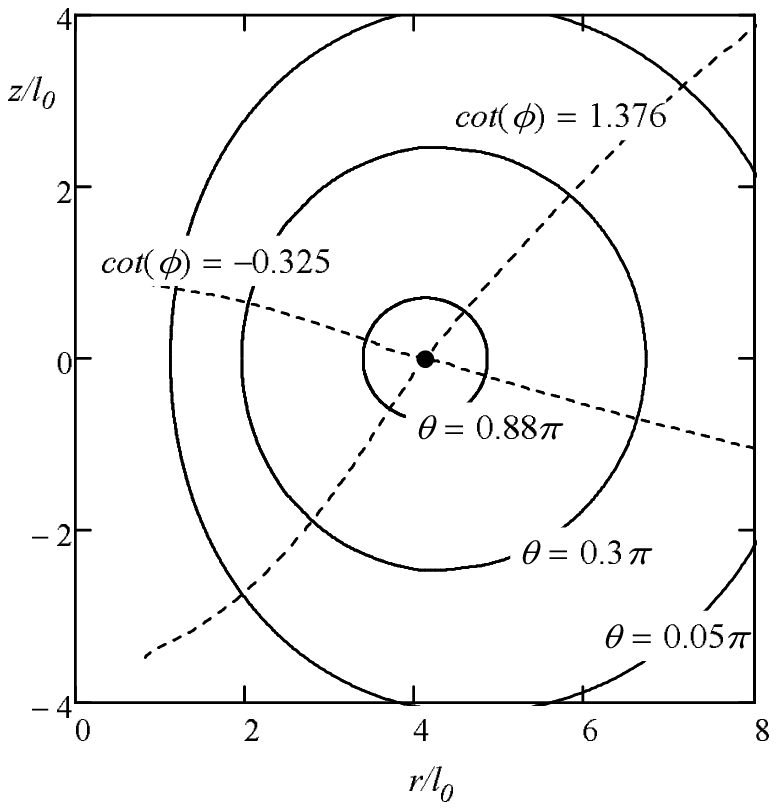}}
\caption{Contours of the angles parameterizing the unit vector ${\bf n}$,
for the (a) stationary hopfion and (b) hopfion moving
with the velocity $V=0.134 V_0$ with the same index $H=3$ and $\omega=0.567\omega_0$. The solid and dashed curves correspond to  $\theta=const$ and $cot(\phi)=const$, respectively.}
\label{F:fig1}
\end{center}
\end{figure}

\begin{figure}
\begin{center}
\includegraphics[width=0.46\columnwidth, viewport=120 380 390 730, clip]{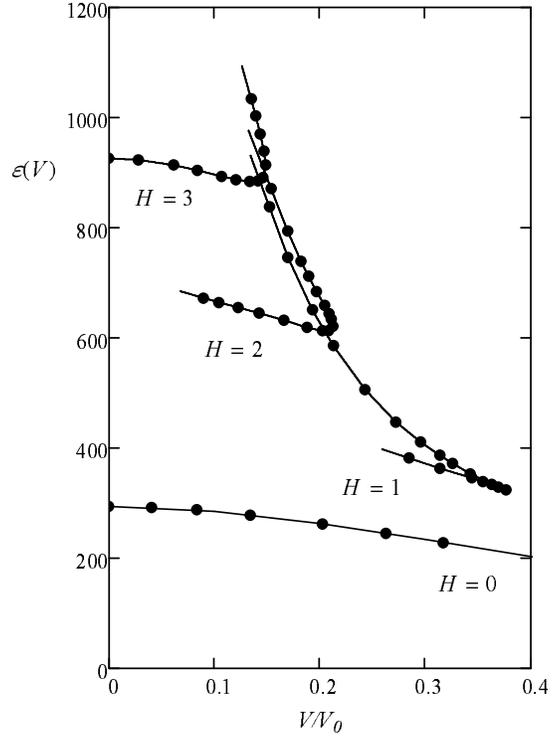}
\caption{Velocity dependence of the energy for nontopological ($H=0$) and topological ($H=1, 2, 3$) solitons with the
same precession frequency $\omega=0.567\omega_0$. The points are
the results of the numerical calculation.}
\label{F:fig2}
\end{center}
\end{figure}

\begin{figure}
\begin{center}
\subfloat[]{\label{F:fig3a}\includegraphics[width=0.46\columnwidth, viewport=120 430 380 690, clip]{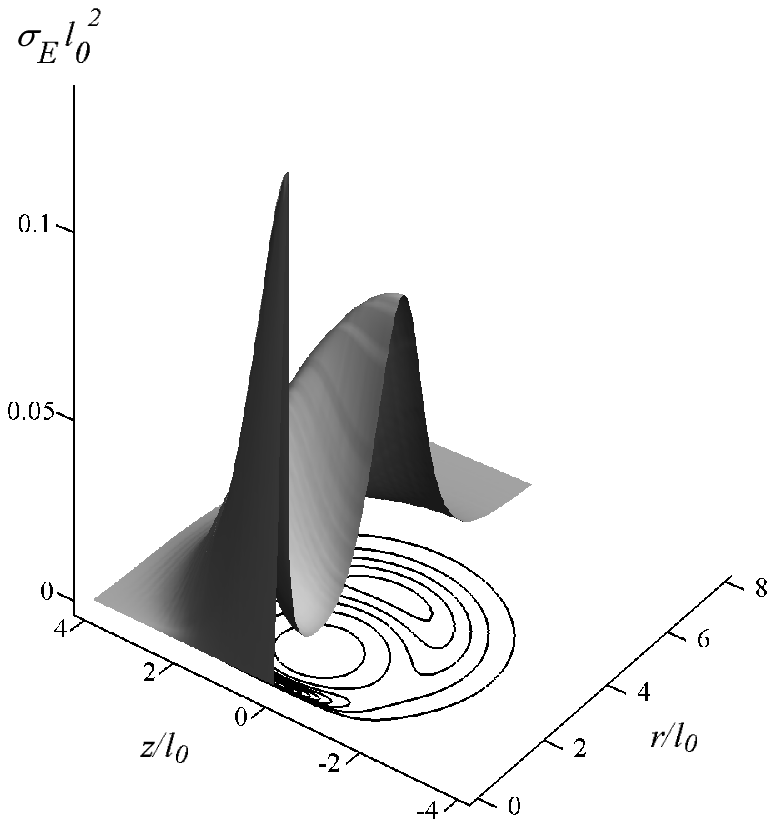}}
\subfloat[]{\label{F:fig3b}\includegraphics[width=0.46\columnwidth, viewport=120 430 380 690, clip]{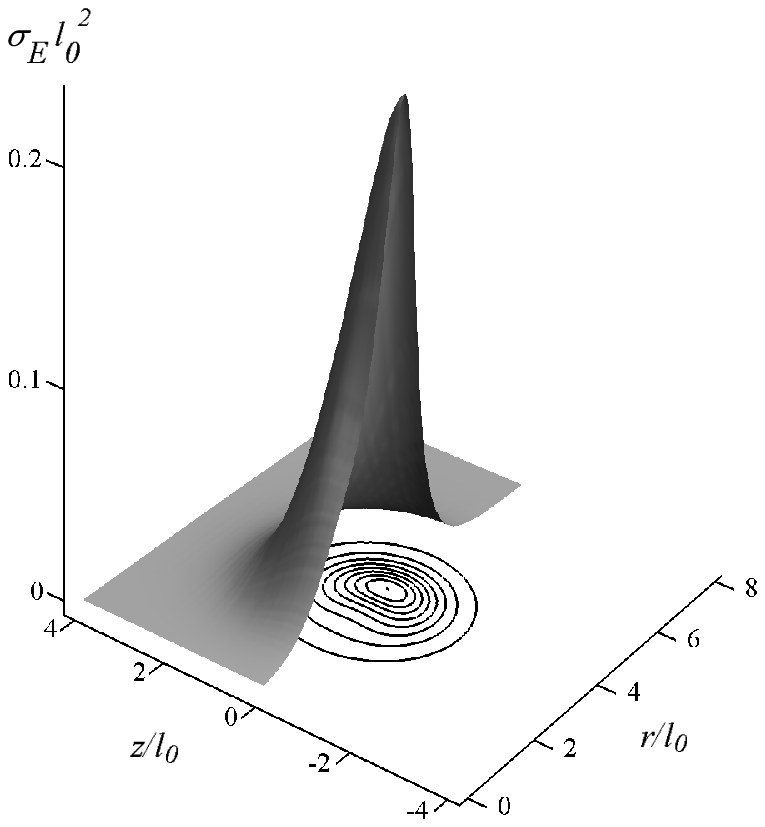}}
\caption{Coordinate dependence of the normalized energy
density for hopfions of the (a) lower and (b) upper energy
branches for $H=1$, $\omega=0.567\omega_0$, $V=0.314{V_0}$.}
\label{F:fig3}
\end{center}
\end{figure}

Fig.\ref{F:fig1} shows the contours of the angles parameterizing vector $\bf{n}$. The coordinates are normalized to
the characteristic length 
\begin{equation}
l_0 = \sqrt{\alpha/\beta}.
\end{equation} 
It is seen that the radius of the central vortex ring corresponding to the value $\theta=\pi$, i.e., to the southern pole
of the ${\mathbb{S}}^2$ sphere, is larger for the moving soliton. The
contours $cot(\phi)=const$ are not constructed near
small $\theta$ values, because this numerical method determining the vectors ${\bf n}$ does not provide an accurate calculation of the azimuth angle when $\mid{n_z}\mid\rightarrow 1$. However, the shapes of these lines are significantly different
for stationary and moving hopfions. The localization
region of the moving soliton is somewhat larger, but is
about $l_0$ as in the stationary case.  

\begin{figure}
\begin{center}
\includegraphics[width=0.46\columnwidth, viewport=100 520 350 715, clip]{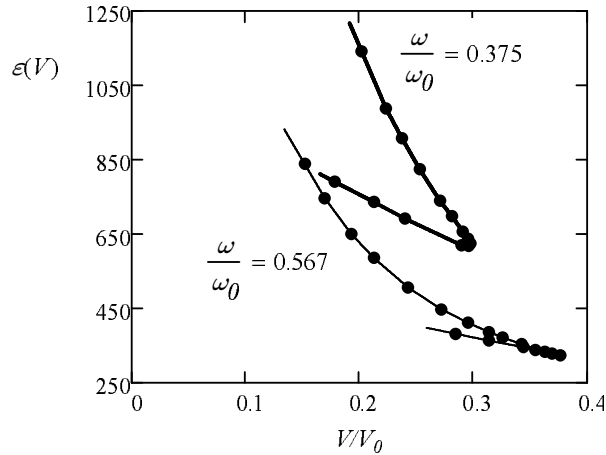}
\caption{Velocity dependence of energy for the hopfion with $H=1$ for two values of the precession frequency.}
\label{F:fig4}
\end{center}
\end{figure}

Fig.\ref{F:fig2} shows the reduced energy $\varepsilon=E/(\alpha {M_0}^2 l_0)$ as
a function of the ratio $V/V_0$, where $V_0=\gamma M_0 \sqrt{\alpha \beta}$. The
plot manifests an important revealed property, the
existence of two types of moving hopfions with the
same $H$, $V$ and $\omega$ values at least in a certain velocity
interval. 
Fig.\ref{F:fig3} presents typical distributions of the
normalized energy density $\sigma_E=w_E/E$. A high energy
density along the wall of the toroidal surface is characteristic of hopfions of the lower energy branch with low
velocities, whereas the energy density increases from
the wall to the center of the toroid for the hopfions of
the upper energy branch with high energies. It is also
seen in Figs. \ref{F:fig2} and \ref{F:fig4}  that the velocity of topological
solitons is limited. The limiting velocity of a hopfion
decreases with decreasing precession frequency,
whereas its energy increases. The frequency $\omega$ is normalized to the frequency of the homogeneous ferromagnetic resonance:
\begin{equation} 
\omega_0=\gamma M_0 \beta.
\end{equation}

The general notions on the structure of the stationary hopfion with $H=1$ cannot allow for a comparison
of its energy with the energy of the corresponding nontopological soliton~\cite{bib:kniga1,bib:KBK}. The features of the structure of the class of objects under investigation prevent
a numerical analysis of the case with $H=1$ and $V=0$~\cite{bib:BorRyb1}. However, the extrapolation of the $\varepsilon(V)$ dependence to the region of small $V$ values in Fig.\ref{F:fig2} for $H=1$ certainly indicates that the energy of the stationary
precession topological soliton is higher than that for
the nontopological soliton at the same frequency $\omega$ at
least in a certain precession frequency range.

\end{document}